\documentclass[preprint]{article}
\newcommand{\be}{\begin{equation}} \newcommand{\ee}{\end{equation}} 
\newcommand{\bea}{\begin{eqnarray}}\newcommand{\eea}{\end{eqnarray}}

\begin{document}
\title{ A note on topological insulator phase in non-hermitian quantum systems}
\author{ Pijush K. Ghosh\footnote{{\bf e-mail:}
pijushkanti.ghosh@visva-bharati.ac.in}}
\date{Department of Physics, Siksha-Bhavana,\\ 
Visva-Bharati University,\\
Santiniketan, PIN 731 235, India.}
\maketitle
\begin{abstract} 
Examples of non-hermitian quantum systems admitting
topological insulator phase are presented in one, two and three space
dimensions. All of these non-hermitian Hamiltonians have entirely real
bulk eigenvalues and unitarity is maintained with the introduction
of appropriate inner-products in the corresponding Hilbert spaces.
The topological invariant characterizing a particular phase is
shown to be identical for a non-hermitian Hamiltonian and its hermitian
counterpart, to which it is related through a non-unitary similarity
transformation. A classification scheme for topological insulator phases in 
pseudo-hermitian quantum systems is suggested.
\end{abstract}
\tableofcontents{}

\section{Introduction}

Topological ideas have resurfaced very often in analyzing physical problems,
the recently discovered topological insulator\cite{kane, qshe_th,
qshe_expt, ti_th,ti_expt,ryu1,ryu2} 
being one such example.
The main characteristic of a topological insulator is the appearance of
boundary zero mode within the bulk gap, the stability of which is guaranteed
due to the existence of some associated topological invariant. The
classification of topological insulators are based on the underlying
discrete symmetries of the Hamiltonian admitting such phases,
like parity(${\cal{P}}$), time-reversal(${\cal{T}}$) and particle-hole symmetry
\cite{ryu1,ryu2}.
Perturbations and/or deformations preserving these symmetries can not
destabilize a phase characterized by a topological invariant, which is
usually an integer or a $Z_2$ quantity, as long as the bulk gap remains open.
Decay of a particular phase within a given topological sector due to the
time-evolution of the states is also ruled out completely for hermitian
quantum systems.

The hermiticity of an operator crucially depends on the metric of the
Hilbert space on which it is defined. In the standard treatment of quantum
mechanics, the metric is always taken to be an identity operator.
An emergent view\cite{bend,ali,quasi,piju_dirac,piju_susy,piju_nr}
in the context of the ${\cal{PT}}$ symmetric and/or pseudo-hermitian quantum
systems is that a consistent non-dissipative description of non-hermitian
quantum systems is admissible with a modified inner-product in the Hilbert
space. A few examples of non-dissipative non-hermitian quantum systems with a complete and consistent description include asymmetric XXZ spin-chain,
non-hermitian transverse Ising model, non-hermitian Dicke model, non-hermitian
quadratic form of bosonic(fermionic) operators and non-hermitian many-particle
rational
Calogero model\cite{piju_nr,piju_misc}. Relativistic\cite{piju_dirac} and
supersymmetric\cite{piju_susy} non-hermitian quantum systems have also been
investigated within the same context. A proposal for optical realization
of non-hermitian relativistic quantum system is described in Ref.
\cite{longhi}. Experimental results related to ${\cal{PT}}$ symmetric optical
systems are also available\cite{expt}.

The purpose of this note is to present a general discussion on the
construction of non-hermitian Dirac Hamiltonians admitting topological
insulator phase, followed by a few explicit examples in one, two
and three space dimensions. The method prescribed by the present Author, in
Refs. \cite{piju_dirac, piju_susy,piju_nr} and particularly in
Ref. \cite{piju_dirac},
is used to construct these models by suitable pseudo-hermitian
deformations of known hermitian Hamiltonians admitting topological
insulator phase.  All the model Hamiltonians presented in this note have
entirely real bulk eigen values and unitary time evolution. 
The topological invariant characterizing a particular topological
insulator phase is identical for the non-hermitian and the corresponding
hermitian Hamiltonian from which it is obtained by pseudo-hermitian
deformation. The decay of a phase within a given topological
sector is also ruled out, since the time-evolution is unitary by construction.

A few attempts have been made in the recent past\cite{hughes,sato} to construct
model non-hermitian Hamiltonians admitting topological insulator phase. The
finding of Ref. \cite{hughes} is that topological insulator phase is absent
in non-hermitian ${\cal{PT}}$ symmetric Hamiltonians. It is worth recalling
in this context that reality of the entire spectra as well as the unitarity
of a non-hermitian Hamiltonian can generally be understood in terms of
an {\it unbroken anti-linear symmetry}, which is not necessarily the standard
${\cal{PT}}$ symmetry\cite{ali,piju_dirac,piju_susy,piju_nr,piju_misc}.
The anti-linear symmetry may be identified as the standard ${\cal{PT}}$
symmetry for some specific quantum systems.
The topological insulator phase in non-hermitian Dirac Hamiltonian having an
anti-linear symmetry is described in this note. The second reference\cite{sato}
contains examples of non-hermitian Hamiltonians admitting topological insulator
phase in which the bulk eigen values of the models are in general complex.
Consequently, the time-evolution is not unitary and the states describing
the topological insulator phase is expected to decay within a given topological
sector. The examples presented in this note are free from these shortcomings;
the entire bulk eigen values are real and the time-evolution is unitary.

The plan of presenting the results is the following. A brief review 
of pseudo-hermitian quantum systems as applied to Dirac Hamiltonians is
presented at the beginning of the next section. Results regarding the allowed
forms of topological invariants for pseudo-hermitian quantum systems are also
included in this section. Examples of non-hermitian Dirac Hamiltonians admitting
topological insulator phase in one, two and three space dimensions are
presented in sections 3.1, 3.2 and 3.3, respectively. Finally, the note ends
with a summary of the results obtained and relevant discussions in section 4.
Suggestions on a possible classification scheme of topological insulators in
non-hermitian quantum systems are included.

\section{Pseudo-hermiticity \& topological invariants}

The continuum description of topological insulators are given in terms of free
particle Dirac Hamiltonians with translational invariance in $D+1$ space-time
dimensions. The main objective of this note is to present Dirac Hamiltonians
$H$, admitting topological insulator phase, which are non-hermitian in the
Hilbert space ${\cal{H}}_D$ that is endowed with the standard inner-product
$\langle \cdot | \cdot \rangle$. Any operator that is related to its adjoint
in ${\cal{H}}_D$ through a non-unitary similarity transformation is known as
pseudo-hermitian operator\cite{ali}, i.e,
\be
H^{\dagger} = \eta \ H \ \eta^{-1}.
\label{pseudo_def}
\ee
\noindent A positive-definite similarity operator $\eta :=\eta_+$, if it
exists, can be identified as a metric operator. A Hilbert space that is
endowed with the metric $\eta_+$ and the inner-product $\langle \langle \cdot|
\cdot \rangle \rangle_{\eta_+}:= \langle \cdot | \eta_+ \cdot \rangle$ is
denoted as ${\cal{H}}_{\eta_+}$. The Hilbert spaces ${\cal{H}}_D$ and
${\cal{H}}_{\eta_+}$ are identical in the limit $\eta_+$ being an identity
operator.

In general, hermitian operators in ${\cal{H}}_{\eta_+}$ are non-hermitian in
${\cal{H}}_D$ and the vice verse. A hermitian Hamiltonian $H$ in
${\cal{H}}_{\eta_+}$ is related to a hermitian Hamiltonian $h$ in
${\cal{H}}_D$ through a non-unitary similarity transformation\cite{ali,quasi},
\be
h = \rho \ H \ \rho^{-1}, \ \ \rho:= \sqrt{\eta_+}.
\label{quasi}
\ee
\noindent An operator $O_{{\eta_+}}$ in ${\cal{H}}_{\eta_+}$ may be introduced
corresponding to each operator $O_{D}$ in ${\cal{H}}_{D}$ as
$O_{D}= \rho O_{\eta_+} \rho^{-1}$ so that
$ \langle O_D \rangle= \langle \langle O_{\eta_+} \rangle \rangle_{\eta_+}$.
This relation allows to find the symmetry generators of $H$ from that of
$h$ or the vice verse and defines physical observables in ${\cal{H}}_{\eta_+}$.
Both the operators $O_{{D}}$ and $O_{\eta_+}$ are hermitian in ${\cal{H}}_D$
as well as in ${\cal{H}}_{\eta_+}$ for the special case for which these
operators commute with the similarity operator $\rho$.

A comment is in order on the nature of the mapping described by Eq.
(\ref{quasi}). The non-hermitian Hamiltonian $H$ may be mapped to a
hermitian Hamiltonian $\tilde{h} = \tilde{\rho} H {\tilde{\rho}}^{-1}$ in
${\cal{H}}_D$ that is different from $h$. Although the existence of such a
similarity operator $\tilde{\rho}$ indicates non-uniqueness of the mapping
(\ref{quasi}), the hermitian Hamiltonians thus obtained are known\cite{ali}
to be unitary equivalent to each other, i.e. $\tilde{h} = U^{-1} h U, \
U:= \rho {\tilde{\rho}}^{-1}$. The unitary nature of $U$ follows\cite{ali}
from the identities $ \rho^{-1} \eta_+ \rho^{-1} = 1 = {\tilde{\rho}}^{-1}
\eta_+ {\tilde{\rho}}^{-1}$. Without loss of any generality, the symmetry
generators of $H$ can thus be obtained by using the inverse-similarity
transformation of  the corresponding generators of either $h$ or $\tilde{h}$.
For example, a symmetry generator $T_D$ of $h$ is related to the corresponding
generator $\tilde{T}_D$ of $\tilde{h}$ though the unitary transformation:
$\tilde{T}_D = U^{-1} T_D U$. Consequently, the symmetry generators
$T_{\eta_+}:= \rho^{-1} T_D \rho$ and ${\tilde{T}}_{\eta_+}:= \rho^{-1}
\tilde{T}_D \rho$ of $H$ are related to each other through a unitary
transformation, $T_{\eta_+} = U^{-1} {\tilde{T}}_{\eta_+} U$ and
$\langle \langle T_{\eta_+} \rangle \rangle_{\eta_+} =
\langle \langle \tilde{T}_{\eta_+} \rangle \rangle_{\eta_+} =
\langle T_D \rangle = \langle \tilde{T}_D \rangle$.
This implies that the expectation values of the physical observables and/or
associated topological invariants of $H$ can be computed either from that
of either $h$ or $\tilde{h}$. The different choices of the similarity operator
$\tilde{\rho}$ correspond to different quantum canonical transformations
in ${\cal{H}}_D$. Thus, without any loss of generality, the discussions below
involve the similarity operator $\rho$.

The Hamiltonians $H$ and $h$ are isospectral. The Bloch eigen states
$|\phi_a(\bf {p}) \rangle$ of $H({\bf p})$ at each $p$ with the eigen values
$E_a({\bf p})$ are related to the corresponding states
$|\psi_a(\bf {p})\rangle$ of $h$ through the equation,
\be
|\phi_a(\bf {p}) \rangle= \rho^{-1} |\psi_a({\bf {p}})\rangle.
\label{state}
\ee
\noindent Further, $|\psi_a({\bf p})\rangle$ constitute a complete set of
orthonormal states in ${\cal{H}}_D$, while the completeness and orthonormality
of states $|\phi_a({\bf p}) \rangle$ can be shown only in ${\cal{H}}_{\eta_+}$.
It is assumed throughout this note that a bulk band gap exists for $h$ that is
centered around some fixed energy(which may be chosen to be zero without loss
of any generality) and the quantum ground state is obtained by filling all the
$N$ states below this level. The isospectrality between $h$ and $H$ allows
to assume identical conditions for $H$. The Berry connection for the
$N$ number of filled Bloch states for $h$ in the Hilbert space
${\cal{H}}_{D}$ is defined as\cite{ryu1,ryu2},
\be
a_{i}^{\hat{a}\hat{b}}({\bf p}) {dp}_{i}:=
\langle \psi_{\hat{a}}({\bf p})|d\psi_{\hat{b}}({\bf p})\rangle,\ \
\ \hat{a}, \hat{b}=1, 2,\dots N, \ \ i=1, 2, \dots D.
\ee
\noindent The Berry connection for the non-hermitian Hamiltonian $H$ is
introduced as,
\bea
{\cal{A}}_i^{\hat{a}\hat{b}}({\bf p}) dp_i
& := & \langle \langle \phi_a({\bf p})|d\phi_b({\bf p})\rangle \rangle
_{\eta_+},\nonumber \\
& = & a_{i}^{\hat{a}\hat{b}}({\bf p}) {dp}_{i}.
\label{berry_pot}
\eea
\noindent Apart from an implicit assumption that $\rho$ is independent of
the momentum ${\bf p}$, Eq.(\ref{state}) and the modified inner-product
in ${\cal{H}}_{\eta_+}$  have been used to obtain the expression in the
second line of the above equation.
The crucial observation at this point is that the gauge potentials
${\cal{A}}_i^{\hat{a}\hat{b}}$ in ${\cal{H}}_{\eta_+}$ and
$a_i^{\hat{a}\hat{b}}$
in ${\cal{H}}_D$ are identical, leading to the same Berry curvature.
Consequently, the Chern numbers and Chern-Simons invariants will have identical
values in ${\cal{H}}_D$ as well as in ${\cal{H}}_{\eta_+}$. The readers are
referred to Ref. \cite{ryu2} for explicit expressions of different topological
invariants in terms of Berry curvature.

The topological invariants associated with chiral topological insulators
are constructed in terms of the projectors onto the filled Bloch states.
In this context, an idempotent operator $\Gamma_D$ may be introduced in odd
space dimensions that anti-commutes with the Dirac Hamiltonian $h$ of a massive
free particle,
\be
\Gamma_D^2=1, \ \ \{h, \Gamma_D \}=0.
\label{chiral_h}
\ee
\noindent The existence of such an operator implies that corresponding to
each eigen state $|\psi_a({\bf {p}})\rangle$ of $h$ with the eigen value $E_a$,
$|\tilde{\psi}_a({\bf {p}})\rangle = \Gamma_D |\psi_a({\bf {p}})\rangle$
is an eigenstate of $h$ with the eigenvalue $-E_a$. 
The operator that anti-commutes with $H$ for a given $\Gamma_D$ may
be found as $\Gamma_{\eta_+}= \rho^{-1} \Gamma_D \rho$. Consequently,
$\Gamma_{\eta_+}$ relates the eigenstates $|\tilde{\phi}_a({\bf p}) \rangle$
and $|\phi_a({\bf p}) \rangle$ of $H$ corresponding to the eigen values
$-E_a$ and $E_a$, respectively. In particular, $|\tilde{\phi}_a({\bf p})
\rangle = \Gamma_{\eta_+} |\phi_a({\bf p}) \rangle$.

The projector $P_{\eta_+}({\bf p})$ onto the filled Bloch states of the
Hamiltonian $H$ at fixed $p$  and the associated ``$Q_{\eta_+}$ matrix"
is defined as,
\bea
P_{\eta_+}({\bf p}) & := & \sum_{\hat{a}} |\phi_{\hat{a}}({\bf p})
\rangle \langle \phi_{\hat{a}}({\bf p})| \eta_+,\nonumber \\
Q_{\eta_+}({\bf p}) & := & 1 - 2 P_{\eta_+}({\bf p}),
\label{pseudo_proj}
\eea
\noindent where $\hat{a}$ indicates a summation over all the filled Bloch
states. The operators $P_{\eta_+}({\bf p})$ and $Q_{\eta_+}({\bf p})$ are
hermitian in the Hilbert space ${\cal{H}}_{\eta_+}$ and satisfy the standard
relations:
\be
P_{\eta_+}^2({\bf p}) = P_{\eta_+}({\bf p}),\ \
Q_{\eta_+}^2({\bf p})=1.
\ee
\noindent The projector $P_D({\bf p})$ onto the filled Bloch states of the
Hamiltonian $h$ at fixed $p$ and the associated ``$Q_D$ matrix" is related
to $P_{\eta_+}({\bf p})$ and $Q_{\eta_+}({\bf p})$ through a non-unitary
similarity transformation,
\bea
&& P_D({\bf p}) := \sum_{\hat{a}}| \psi_{\hat{a}} \rangle \langle
\psi_{\hat{a}}|= \rho \ \ P_{\eta_+}({\bf p}) \ \rho^{-1},\nonumber \\
&& Q_D({\bf p}) := 1 - 2 P_D({\bf p}) = \rho \ Q_{\eta_+}({\bf p}) \rho^{-1},
\label{topo_eq}
\eea
\noindent where $\hat{a}$ indicates a summation over all the filled Bloch
states and $P_D^2({\bf p})=P_D({\bf p}), Q_D^2({\bf p})=1$.
The topological invariants for class $AIII$ topological insulators
in hermitian quantum systems is constructed by using the projector
$Q_D({\bf p})$\cite{ryu1,ryu2} and Eq. (\ref{topo_eq}) implies that
topological invariants for $H$ and $h$ are identical.

A comment is in order before the end of this section. The existence of
$\Gamma_D$ implies that the pseudo-hermitian $H$ is also
pseudo-anti-hermitian\cite{sato} with respect to an operator $\kappa$:
\be
H^{\dagger} = - \kappa H \kappa^{-1}, \ \
\kappa := \rho \ \Gamma_D \rho = \eta_+ \Gamma_{\eta_+}.
\ee
\noindent The types of the operators $\Gamma_D$ and $\rho$ those will be
considered in this note satisfy the identity,
\be
\rho \ \Gamma_D \ \rho = \Gamma_D.
\label{anti-h}
\ee
\noindent The hermitian Hamiltonian $\tilde{H}:=H+H^{\dagger}$ in ${\cal{H}}_D$
anti-commutes with the operator $\Gamma_D$, when the validity of Eq.
(\ref{anti-h}) is assumed, implying that $\tilde{H}$ can be transformed into a
block off-diagonal form in a representation
in which $\Gamma_D$ is diagonal. The projector onto the filled
Bloch states of the Hamiltonian $\tilde{H}$ can be used to construct
topological invariants corresponding to $H$ provided $H$ is a normal
operator, i.e. $[H,H^{\dagger}]=0$, so that $\tilde{H}$ and $H$ are
simultaneously diagonalizable. The non-hermitian Hamiltonians considered
in this note are not normal operators and hence, this scheme of constructing
topological invariants is not applicable. On the other hand, the projector
$Q:= 1 -\left [ P_{\eta_+}({\bf p}) + P_{\eta_+}^{\dagger}({\bf p}) \right ]$
onto the filled Bloch states of $H$ and $H^{\dagger}$ is not an idempotent
operator. Consequently, a block in the block off-diagonal form of $Q$ can
not necessarily be identified as an element of $U(N)$. Thus, the use of $Q$
to construct topological invariant corresponding to non-hermitian $H$ seems
problematic.

\section{Examples}

A few examples of non-hermitian Hamiltonians admitting topological insulator
phase are presented in this section. All the non-hermitian Hamiltonians
considered in this note can be mapped to hermitian Hamiltonians, through
non-unitary similarity transformations, which are known to admit topological
insulator phase. It follows from the discussions in the previous
section that the bulk eigen values and the topological invariants are identical
for both the non-hermitian Hamiltonian and its similarity transformed hermitian
version. The bulk eigen states, zero modes and symmetry generators of the
non-hermitian Hamiltonian can be obtained from the corresponding quantities
of the hermitian Hamiltonian through the inverse similarity transformation.
Thus, in general, only non-hermitian Hamiltonians and associated non-unitary
similarity transformations are mentioned explicitly in this note to avoid any
repetition of known results. The $1+1$ dimensional Dirac equation is treated
in some detail in the next section to give a general outline of the technique
involved that can be applied under similar situations.

\subsection{ Dirac Hamiltonian in $1+1$ dimensions}
The first example considered in this note is the $1+1$ dimensional
non-hermitian Dirac Hamiltonian,
\be
H^{(1)} = \sigma^2 p_x + m(x) \ cosh \phi \ \sigma^3 
- i \ m(x) \ sinh \phi \ \sigma^1, \ \phi \in R,
\ee
\noindent where $p_x$ is the linear momentum and $m(x)$ is an arbitrary
real function of its argument. The Hamiltonian $H^{(1)}$ was first
introduced \footnote{The Dirac Hamiltonian $H_{1D}$ in Eq.(6) of
Ref. \cite{piju_dirac} is reduced to $H^{(1)}$ for $M(x)=V(x)=0$ and
$P(x)=-m(x)$. } in Ref. \cite{piju_dirac} in the context of non-dissipative
non-hermitian relativistic quantum system in ${\cal{H}}_D$. It was
shown\cite{piju_dirac} that $H^{(1)}$ is hermitian in
the Hilbert space ${\cal{H}}_{\eta_+^{(1)}}$ that is endowed with the metric
$\eta_+^{(1)}$,
\be
\eta_+^{(1)} := e^{-\phi \sigma^2}, \ \
\rho^{(1)} := e^{-\frac{\phi}{2} \sigma^2}.
\ee
\noindent The non-unitary similarity operator $\rho^{(1)}$ maps $H^{(1)}$ to 
a hermitian Hamiltonian $h^{(1)}$ in ${\cal{H}}_D$\cite{piju_dirac},
\be
h^{(1)}:= \rho^{(1)}  H^{(1)} (\rho^{(1)})^{-1}= \sigma^2 p_x + m(x) \sigma^3,
\label{sim1}
\ee
\noindent implying that $H^{(1)}$ and $h^{(1)}$ are isospectral. The
Hamiltonian $h^{(1)}$ anti-commutes with $\sigma^1$. Correspondingly, the
Hamiltonian
$H^{(1)}$ anti-commutes with the operator $(\rho^1)^{-1} \sigma^1 \rho^{(1)} =
\sigma^1 \ \eta_+^{(1)}$.

The Dirac Hamiltonian $h^{(1)}$ with $m(x)=m \in R$ provides an example of
a chiral topological insulator in class $AIII$\cite{ryu2}. The relation
(\ref{sim1}) implies that the non-hermitian Hamiltonian $H^{(1)}$ also admits
topological insulator phase. The bulk spectrum of $H^{(1)}$ contains
a mass gap, $E_{\pm}=\pm \sqrt{p_x^2 + m^2} \equiv \pm \lambda$ and the
Bloch wave function corresponding to the negative eigenvalue state reads,
\be
|\phi^-(p_x)\rangle = \frac{(\rho^{(1)})^{-1}}{2 \sqrt{p_x^2 + m^2}}
\pmatrix {{ip_x - m + \lambda}\cr \\ {-(ip_x-m) +\lambda}}.
\ee
\noindent The Berry connection ${\cal{A}}(p_x)$ and the associated
Chern-Simons invariant $CS_1$ for the state $|\phi^-(p_x)\rangle$ may
be computed as,
\be
{\cal{A}}(p_x)= - \frac{i m}{2 \lambda^2} dp_x, \ \
CS_1=\frac{m}{4{\mid m \mid}}, 
\ee
\noindent which are identical with the corresponding expressions\cite{ryu2} for
$h^{(1)}$ in ${\cal{H}}_D$. The projector $Q_{\eta_+}$ can be expressed
as follows,
\be
Q_{\eta_+} = \left (U \rho^{(1)} \right )^{-1} \ \pmatrix { {0} & {q}\cr \\
{q}^* & {0}} \ \left ( U \rho^{(1)} \right ), \ \
q:= \frac{-i p_x + m}{\lambda}, \ \ U:= e^{-\frac{i \pi}{4} \sigma^2},
\ee
\noindent where $q^*$ denotes the complex conjugate of $q$. It may be
noted that the unitary operator $U$ and the similarity operator $\rho^{(1)}$
are independent of $p_x$ and $m$. The winding
number is calculated as, $\nu=\frac{i}{2 \pi} \int q^{-1} dq = \frac{m}{2
{\mid m \mid}}$, which is twice the value of the Chern-Simons invariant $CS_1$. 

The operator $h^{(1)}$ can be identified as the real supercharge
of a ${\cal{N}}=1$ supersymmetric quantum system. The ground state of the
supersymmetric Hamiltonian $h_s:= [h^{(1)}]^2$ in the supersymmetry-preserving
phase is a zero mode of $h^{(1)}$. The zero mode of $H$ can be obtained from
the zero mode of $h^{(1)}$ by using Eq. (\ref{state}). For example, assuming
that $Lt_{x \rightarrow \pm \infty} m(x) = \pm m, m \in R^+$, the zero mode
of $H$ reads,
\be
\phi_0 = \frac{1}{\sqrt{2}} \left ( \rho^{(1)} \right )^{-1} \
e^{-\int_0^x m(x) dx} \pmatrix{{1}\cr \\{-1}} .
\ee
\noindent The zero mode of $H$ for different shapes of $m(x)$ may be obtained
in a similar way from the corresponding well-behaved zero-energy state
of $h^{(1)}$.

\subsection{ Dirac Hamiltonian in $2+1$ dimensions}

The second example is a $4$-component non-hermitian Dirac equation in $2+1$
dimensions. The relevant gamma matrices may be constructed in terms of the
elements of the Clifford algebra,
\be
\{ \xi^p, \xi^q \} = \delta^{pq}, \ \ p, q=1, 2, \dots, 5.
\ee
\noindent The three generators of the group $O(3)$ is expressed as,
\be
J^a := \frac{i}{8} \epsilon^{abc} \left [ \xi^b, \xi^c \right ], \ \
a, b, c=1, 2, 3.
\ee
\noindent The Hilbert space ${\cal{H}}_{\eta_+^{(2)}}$ is endowed with the
metric $\eta_+^{(2)}$,
\be
\eta_+^{(2)}:= e^{-\phi \ \hat{n} \cdot \vec{J} }, \ \
\rho^{(2)} := e^{-\frac{\phi}{2} \hat{n} \cdot \vec{J}}, \ \
\phi \in R, \ \ \hat{n} \cdot \hat{n} =1.
\ee
\noindent The Hamiltonian $H^{(2)}$,
\bea
&& H^{(2)} = \xi^4 p_x + \xi^5 p_y + m \sum_{b=1}^3 \ R^{3b} \xi^b,\nonumber \\
&& R^{ab} \equiv n^a n^b ( 1 - cosh \phi ) + \delta^{ab} cosh \phi
+ i \epsilon^{abc} n^c sinh \phi, 
\eea
\noindent is non-hermitian in ${\cal{H}}_D$ and hermitian in
${\cal{H}}_{\eta_+^{(2)}}$. The non-hermiticity arises in $H^{(2)}$ arises
due to the fact that the complex conjugate of $R^{ab}$ is not equal to itself.

The non-unitary similarity operator $\rho^{(2)}$ maps $H^{(2)}$ to a hermitian
Hamiltonian $h^{(2)}$ in ${\cal{H}}_D$,
\bea
h^{(2)} & := & \rho^{(2)} H^{(2)} (\rho^{(2)})^{-1}\nonumber \\
& = & \xi^4 p_x + \xi^5 p_y + m \xi^3.
\eea
\noindent The elements of the Clifford algebra are realized in terms of the
following matrices,
\be
\xi_1:= \tau^2 \otimes I, \ \ \xi_2:= \tau^3 \otimes I, \ \
\xi_3:= \tau^1 \otimes \sigma^3, \ \ \xi_4:= \tau^1 \otimes \sigma^1, \ \
\xi_5:= \tau^1 \otimes \sigma^2,
\label{rep_cliff}
\ee
\noindent where $\tau^a, \sigma^a$ with $a=1,2,3$ are two sets of Pauli matrices
corresponding to two different sub-lattices and $I$ is a $2 \times 2$ identity
matrix. The Hamiltonian $h^{(2)}$ is known to admit topological insulator
phase\cite{ryu2}.
In fact, $h^{(2)}$ describes two copies of the two dimensional two-component
Dirac Hamiltonian having topological insulator phase. Following the general
discussions in section 2, the conclusion is that $H^{(2)}$ admits topological
insulator phase with the identical bulk eigenvalues and topological invariant
as that of Hamiltonian $h^{(2)}$. The bulk eigen states as well as the zero
mode state of $H^{(2)}$ can be obtained from the corresponding states of
$h^{(2)}$ by using the relation ( \ref{state}) and identifying
$\rho:=\rho^{(2)}$.

\subsection{ Dirac Hamiltonian in $3+1$ dimensions}
A Hamiltonian in $3+1$ dimensions may be introduced as follows,
\bea
&& H^{(3)} = \vec{\alpha} \cdot \vec{p} + m \ e^{\gamma_5 \phi} \beta,
\ \ \phi \in R\nonumber \\
&& \alpha^a := \tau^1 \otimes \sigma^a, \ \ 
\beta := \tau^3 \otimes I, \ \
\gamma^5 := \tau^1 \otimes I, \ \ a=1, 2, 3,
\eea
\noindent where the Dirac-representation of  the $\gamma$ matrices has
been used and $\tau^a, \sigma^a$ correspond to two independent sets of Pauli
matrices. The Hamiltonian is non-hermitian in the Hilbert space ${\cal{H}}_D$
for $\phi \neq 0$. The generator of the time-reversal transformation is
defined as,
${\cal{T}}:= \left ( I \otimes i \sigma^2 \right ) K, 
$
where $K$ is the complex conjugation operator.
The Hamiltonian $H^{(3)}$ is invariant under the time-reversal transformation.
The last term spoils the invariance of the Hamiltonian under the
parity transformation with the standard form of the generator
${\cal{P}}:=\beta$. However, a non-standard generator of the parity
transformation may be introduced as $\tilde{\cal{P}}:=
e^{\phi \gamma_5} \beta$, which keeps $H^{(3)}$ invariant.

The Hamiltonian $H^{(3)}$ becomes hermitian in ${\cal{H}}_D$ in the limit
$\phi \rightarrow 0$ and is hermitian for any arbitrary $\phi$ in the Hilbert
space ${\cal{H}}_{\eta_+}$ that is endowed with a positive-definite metric
$\eta_+^{(3)}$,
\be
\eta_+^{(3)}:= e^{- \phi \ \gamma_5}, \ \
\rho^{(3)}:= e^{-\frac{\phi}{2} \gamma_5}.
\ee
\noindent In fact, the {\it non-unitary similarity operator}
$\rho^{(3)}$ can be used to map $H^{(3)}$ to a hermitian Hamiltonian
$h^{(3)}$,
\bea
h^{(3)} & = & \rho^{(3)} \ H^{(3)} \ (\rho^{(3)})^{-1} \nonumber \\
& = & \vec{\alpha} \cdot \vec{p} + m \ \beta,
\eea
\noindent that is hermitian in ${\cal{H}}_D$. Thus, $H^{(3)}$ and $h^{(3)}$ are
isospectral. It is known that the Hamiltonian $h^{(3)}$ admits topological
insulator phase\cite{ryu2}. Thus, the Hamiltonian $H^{(3)}$ also admits
topological insulator phase with the identical bulk eigenvalues and topological
invariant as that of Hamiltonian $h^{(3)}$. Other relevant quantities of
$H^{(3)}$ may be obtained from $h^{(3)}$ by using Eq. (\ref{state}) with the
replacement of $\rho$ by $\rho^{(3)}$.

\section{Summary \& Discussions}

Examples of non-hermitian Dirac Hamiltonians admitting topological insulator
phase have been presented in one, two and three spatial dimensions. All
these Hamiltonians are hermitian in a Hilbert space that is endowed with a
positive-definite metric and a modified inner-product. The entire
bulk eigenvalues of a given non-hermitian Hamiltonian are thus real. It has
been also shown that any non-hermitian Hamiltonian belonging to this class
can be mapped to a hermitian Hamiltonian through a non-unitary similarity 
transformation, implying that these two quantum systems are isospectral to each
other. Further, the time-evolution of the bulk states as well as the zero modes
of the non-hermitian Hamiltonians are unitary in the Hilbert space
$H_{\eta_+}$. It appears that topological insulator phase in non-hermitian
Dirac Hamiltonians admitting entirely real spectra and unitary time-evolution
have been discussed in this note for the first time in the literature.

The topological insulator phases in hermitian quantum systems have been
classified previously\cite{ryu1,ryu2} according to certain discrete
symmetries of Dirac Hamiltonians. Perturbations preserving the symmetries of
the original Hamiltonian can not destabilize the topological insulator phase
as long as the gap is not closed.
A direct application of this classification scheme to the non-hermitian
Hamiltonians presented in this note may give conflicting results, since
the non-hermitian terms may or may not preserve the symmetry of the hermitian
part of the same Hamiltonian. However, the same classification scheme can be
used for a non-hermitian Hamiltonian provided the identification of the
topological class is based on the analysis of the hermitian Hamiltonian
to which it is related through a non-unitary similarity transformation. It
may be worth emphasizing here that the topological index characterizing a
particular type of insulator, as presented in this note, is identical for
the non-hermitian and the corresponding hermitian Hamiltonians. The
non-hermitian Hamiltonian and its hermitian counterpart thus belong to the
same topological class by construction. It may be emphasized here that the
unitary equivalence between different hermitian Hamiltonians $h, \tilde{h}$
in ${\cal{H}}_D$, those are obtained from the same non-hermitian Hamiltonian
$H$, allows a unique characterization of the topological class of $H$ admitting
topological insulator phases.

The approach taken in this note in classifying topological insulator phases
in non-hermitian quantum systems may equally be extended to any generic
non-hermitian Hamiltonian that can be mapped to a hermitian Hamiltonian.
In general, finding the non-unitary similarity operator that maps the
non-hermitian Hamiltonian to a hermitian one, is a highly non-trivial task.
However, the continuum description of topological insulators are in terms of
Dirac Hamiltonians which involve gamma matrices. The method described
in Ref. \cite{piju_susy} for a pseudo-hermitian realization of the gamma
matrices of arbitrary dimensions may be useful for finding the hermitian
Hamiltonian corresponding to a given non-hermitian Hamiltonian. It is desirable
that the method presented in this note and in previous works\cite{piju_dirac,
piju_susy,piju_nr} is utilized fully to construct at least one physically
realizable non-hermitian Dirac Hamiltonian admitting topological insulator
phase.

.
\end{document}